%% file: paper.tex
\documentclass[10pt,letterpaper]{article}
\usepackage[margin=.85in]{geometry}
\usepackage[T1]{fontenc}
\usepackage{lmodern,microtype,amsmath,amssymb,bm,amsthm}
\usepackage{graphicx,booktabs,array,xcolor,caption,placeins,xurl}
\usepackage{subcaption}
\usepackage[numbers,sort&compress]{natbib}
\usepackage[pdfusetitle,colorlinks=true,linkcolor=blue!45!black,citecolor=blue!45!black,urlcolor=blue!45!black]{hyperref}
\newcommand*{\doi}[1]{doi:~\href{https://doi.org/#1}{{\urlstyle{rm}\nolinkurl{#1}}}}
\graphicspath{{figures/}}
\newcommand{\norm}[1]{\left\lVert #1\right\rVert}
\newcommand{\diag}{\operatorname{diag}}
\newtheorem{proposition}{Proposition}

\input{generated/classical_values}

\input{generated/element_range_values}
\input{generated/p2_read_voltage_values}
\input{generated/p2_generalization_values}

\makeatletter\let\tableinput\@@input\makeatother
\title{A Repeating Analog Finite-Element Circuit Block for\\
3D Continuous and Discontinuous Galerkin Methods}
\author{Arash Ghasemi\\
\small Sumer Analog LLC, 9111 Cross Park Dr., Suite D200, No. 1000, Knoxville, TN 37923, USA\\
\small University of Tennessee at Chattanooga, Chattanooga, TN 37403, USA\\
\small \href{mailto:syn535@tennessee.edu}{syn535@tennessee.edu}}
\date{November 26, 2026}
\begin{document}
\AddToHookNext{env/center/begin}{\vspace{-26pt}}
\maketitle
\begin{abstract}
Finite element assembly maps naturally onto wiring. We present a repeating
analog circuit block for a tetrahedral element that stores the element's
stiffness coefficients and gives every element row its own buffer. A digital
programmer sets the coefficients and switches before the solve, so the same
blocks assemble either continuous Galerkin (CG) or symmetric interior-penalty
discontinuous Galerkin (SIPG) equations without a general-purpose
field-programmable analog array. We derive the scaling that lets each element
fit its coefficients to the available conductance range while preserving both
Galerkin systems. In
LTspice simulations of scalar diffusion, refined linear (P1) and quadratic
(P2) CG circuits with ideal coefficient conductances reproduce digital finite
element solutions to within \PoneCircuitMicroV{} and \PtwoCircuitMicroV{}~$\mu$V
root mean square. With fixed-state memristor reads and transistor buffers, a
25~mV drive chosen on one family of meshes holds the circuit error to
\GeneralizationMinRatioPercent--\GeneralizationMaxRatioPercent\% of the digital
discretization error on six new P2 CG and SIPG cases, well inside a 10\%
budget. For three-dimensional elasticity of a base-clamped Stanford bunny,
simulated with exact coefficient resistors and transistor buffers,
reprogramming the voltage polarities and element current scales lowers the
finest P1 CG circuit error from 146\% to 3.98\% of the discretization error
estimated from a finer P2 solution, and a local tracking-current correction
recovers the digital SIPG solution when its gains are exact.
\end{abstract}

\section{Introduction}
The finite element method (FEM) builds a global system from two ingredients,
a stiffness matrix for each element and a record of the nodes that
neighboring elements share. An analog circuit can be configured from exactly
the same information. The element matrix sets the strengths and signs of the
coefficient branches inside a circuit block, and the mesh connectivity sets
the switches between blocks. Once a digital programmer has configured the
network, its equilibrium voltages represent the discrete solution. We ask how
closely such a circuit, built from devices with a finite conductance range and
imperfect voltage followers, can reproduce the field that digital FEM computes
on the same mesh.

Electrical analogues of partial differential equations have a long history.
They include networks with variable cell size \citep{macneal1949}, passive
circuits for three-dimensional elasticity \citep{barnoski1966}, and circuits
assembled from individual finite element matrices \citep{save2011}. More
recently, memristor arrays have solved PDEs with software-assisted precision
refinement \citep{zidan2018}, other solvers have combined analog iteration with
digital refinement \citep{li2025iteration}, and resistive-memory chips have
performed matrix inversion \citep{zuo2025}. We study a different design
question. If every element scales its coefficients to fit the devices, which
connections still produce the Galerkin equations, and how much error do the
devices add compared with the discretization error that the mesh already
carries?

\begin{figure}[!htbp]\centering
\includegraphics[width=.99\linewidth]{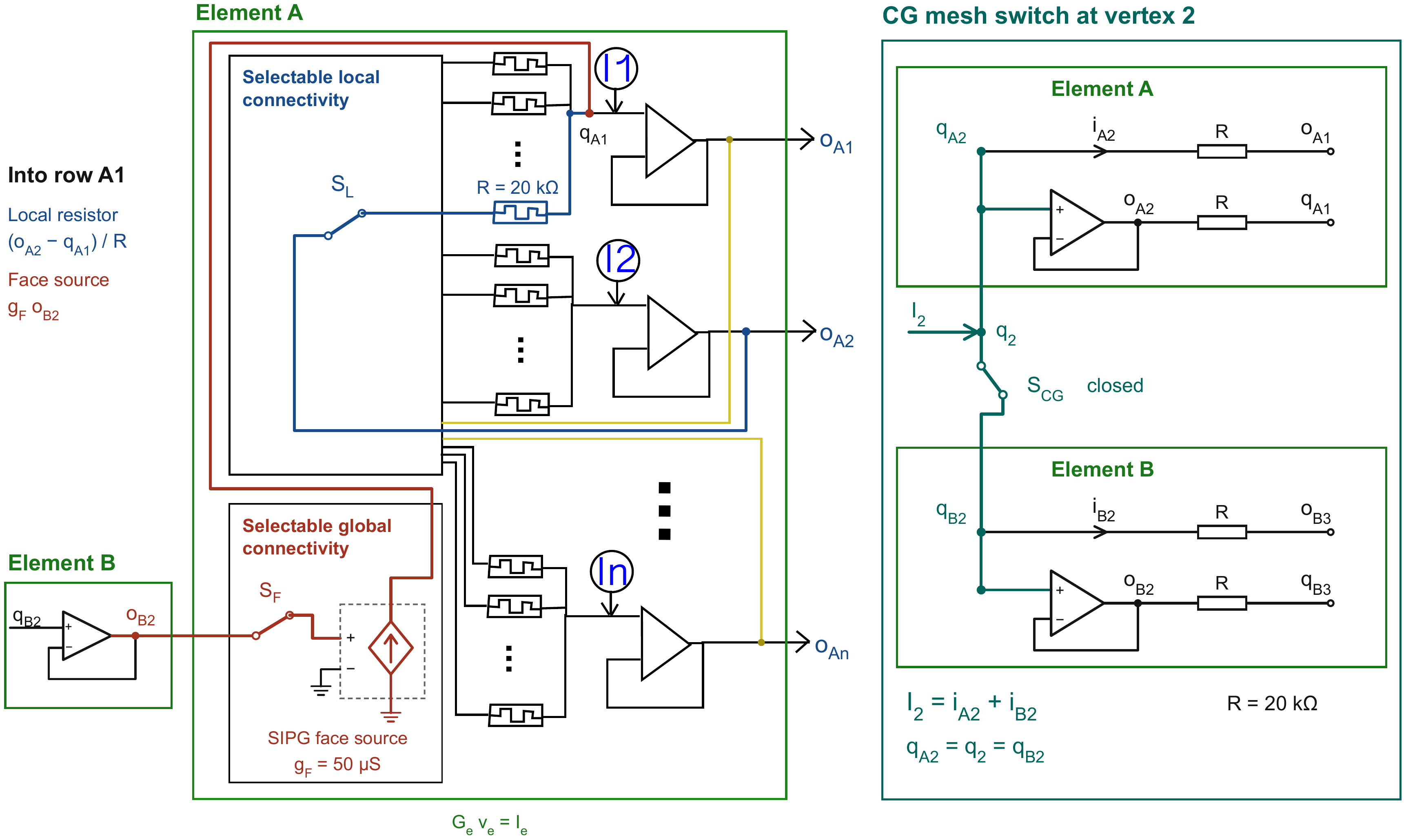}
\caption{Repeating element block with selectable local feedback, a SIPG face connection and a shared CG row port at vertex~2 (inset).}
\label{fig:paper-element}
\end{figure}

Figure~\ref{fig:paper-element} shows the two ways the same block can be
connected. Switch $S_L$ routes a local buffer output through a coefficient
resistor. In SIPG, $S_F$ selects an output of a neighboring element to control
a face current into row $q_{A1}$. In CG, closing $S_{\mathrm{CG}}$ joins the
row ports of two elements at vertex~2, so their currents add at that node
while each buffer still drives only its own element. Each triangle in the
figure is a servo buffer connected as a voltage follower.

Section~\ref{sec:mapping} derives the circuit mapping that preserves CG and
SIPG assembly when every element uses its own scale, and works through a
two-triangle example. Section~\ref{sec:budget} defines an accuracy budget that
compares the circuit error with the discretization error of the same mesh, and
introduces the memristor read law and the row buffer.
Section~\ref{sec:controlled} selects a read voltage on one family of cube
meshes and tests it on another. Section~\ref{sec:refinement} refines meshes
against exact and fine-mesh references, first for scalar diffusion and then
for three-dimensional elasticity. Sections~\ref{sec:operating}
and~\ref{sec:conclusion} discuss what limits the operating range and draw
conclusions.

\section{Assembly-preserving circuit mapping}\label{sec:mapping}
\subsection{Local elements and shared CG nodes}
For the diffusion problem $-\Delta u=f$, let $T_e$ be a straight-sided
tetrahedron. The linear (P1) basis consists of its four barycentric
coordinates $\lambda_i$. The quadratic (P2) basis has four vertex functions
$\lambda_i(2\lambda_i-1)$ and six edge functions $4\lambda_i\lambda_j$. For
either basis, the element stiffness matrix and load vector are
\begin{equation}
 (K_e)_{ij}=\int_{T_e}\nabla N_i\cdot\nabla N_j\,dV,\qquad
 (b_e)_i=\int_{T_e}N_i f\,dV .
\end{equation}
Because a constant field has zero gradient, $K_e\bm1=0$, and the field is
determined only after the elements are assembled and boundary conditions are
applied. If $R_e$ selects the entries of the global voltage vector $u$ that
belong to element $e$, classical CG assembly gives
\begin{equation}
 K=\sum_e R_e^TK_eR_e,\qquad b=\sum_e R_e^Tb_e .
 \label{eq:paper-assembly}
\end{equation}

In the circuit, each element has one row port $q_i$ and one buffered output
$o_i$ for every basis function. Row ports that represent the same CG node share
a voltage, while each buffer drives only the coefficient branches of its own
element. A common factor $\gamma$ converts stiffness coefficients into
conductances, and an element gain $s_e>0$ raises the element's largest
off-diagonal coefficient to the upper conductance limit. In the ideal linear
circuit, the scaled element residual current is
$j_e=\gamma s_e(K_eR_eu-b_e)$. If each element's current is divided by its own
gain before the currents meet at a shared node, then
\begin{equation}
 \sum_e R_e^T\frac{j_e}{s_e}=\gamma(Ku-b),
 \label{eq:paper-decoding}
\end{equation}
and Kirchhoff's current law imposes the assembled FEM equation at every free
node. The CG netlist therefore divides the stiffness currents by $s_e$ and
injects the assembled load $\gamma b$ directly at the shared ports. Under a
uniform change of length at fixed tetrahedron shape, $K_e$ scales in
proportion to length and $s_e$ in inverse proportion. For a change of shape we
recompute $K_e$ from the element Jacobian.

\subsection{Independent DG nodes and face currents}\label{sec:dg-faces}
In SIPG, each element keeps its own nodal values. Consider a face $F$ shared by
a left and a right tetrahedron. We take the unit normal $n$ from left to right,
and write $[u]=u_L-u_R$ for the jump and
$\{\partial_n u\}=(\nabla u_L+\nabla u_R)\cdot n/2$ for the average normal
derivative. For a test function $w$, the face contributes
\begin{equation}
 \int_F\left(-\{\partial_n u\}[w]-\{\partial_n w\}[u]
              +\tau_F[u][w]\right)dS,\qquad
 \tau_F=\frac{4(p+1)^2}{\min(h_{L,F},h_{R,F})}.
 \label{eq:paper-sipg-face}
\end{equation}
Here $p$ is the polynomial degree and $h_{e,F}=3|T_e|/|F|$ is the altitude of
element $e$ over the face. The first two terms couple the normal derivatives
and traces on the two sides, and the third penalizes the jump in voltage. On a
Dirichlet boundary face we use the one-sided symmetric Nitsche terms with the
local altitude and the prescribed boundary value
\citep{ngsolvedg,dealiisips}.

Independent nodal values also let each element use its own voltage scale. Let
$u_e=D_e z_e$, where $z_e$ holds the local circuit coordinates, and choose
\begin{equation}
 (D_e)_{ii}=\sqrt{\frac{\max_j(K_e)_{jj}}{(K_e)_{ii}}},\qquad
 s_e=\frac{G_{\max}}{\gamma\max_{i\ne j}|(D_eK_eD_e)_{ij}|},
 \label{eq:paper-dg-scales}
\end{equation}
so that the diagonal entries of $D_eK_eD_e$ are all equal. Let $A_{ef}$ be the
block of the full SIPG matrix with rows in element $e$ and columns in element
$f$, and let $b_e$ include the volume and boundary loads. The ideal row
equations become
\begin{equation}
 \sum_f\gamma s_e D_e A_{ef}D_f z_f=\gamma s_e D_e b_e .
 \label{eq:paper-dg-transform}
\end{equation}
The sending element supplies $D_f$, and the receiving element supplies
$s_eD_e$, which also multiplies its load. Dividing out these row factors and
recovering $u_f=D_fz_f$ returns the original SIPG system. With
$D=\operatorname{blockdiag}(D_e)$, the congruence $DAD$ is positive definite
whenever $A$ is, although different row gains can make the programmed matrix
nonsymmetric.

\input{paper_two_triangle}

\subsection{Signed coefficients and finite conductance range}
For the scalar diffusion circuits, let $C_e=\gamma s_eD_eK_eD_e$, with $D_e=I$
for CG. A passive branch from $q_i$ to $o_j$ draws a current proportional to
$q_i-o_j$, so its coefficient on $o_j$ is always negative. To realize either
sign, we program a positive conductance $g_{ij}$ and add controlled currents
$h_{ij}o_j$ and $d_io_i$ to the row. For the active off-diagonal entries we use
\begin{equation}
 g_{ij}=\operatorname{clip}(|C_{e,ij}|,G_{\min},G_{\max}),\qquad
 h_{ij}=2\max(C_{e,ij},0),\qquad
 d_i=C_{e,ii}-\sum_{j\ne i}|C_{e,ij}|,
 \label{eq:paper-signed}
\end{equation}
where the sum runs over the active entries in row $i$. The programmer omits
structural zeros and entries at the level of round-off. With ideal followers
and unclipped linear conductances, each off-diagonal coefficient is
$-|C_{e,ij}|+h_{ij}=C_{e,ij}$, and the current $d_io_i$ restores the specified
diagonal. Together the branches reproduce the retained matrix.

The DG face circuit uses signed current sources and diagonal branches. If
$H=A-\operatorname{blockdiag}(K_e)$, its transformed matrix is
$F=\gamma\operatorname{blockdiag}(s_eD_e)HD$. Diagonal terms sense the
receiving port $q_i$, whereas off-diagonal terms sense the sending output
$o_j$. The actual face current is therefore $Fo+\diag(F)(q-o)$, where
$\diag(F)$ keeps only the diagonal of $F$. The second term is the route by
which a follower's tracking error enters the face current, and it reappears in
the refinement studies of Section~\ref{sec:refinement}.

A single element gain can represent all active coefficient magnitudes when
their largest-to-smallest ratio fits within $G_{\max}/G_{\min}$. A second
conductance bank can extend this range by multiplying the small coefficients
by a further factor of 512 and dividing their current by $512s_e$. On a
745-tetrahedron P2 CG mesh with the original servo, ngspice simulations gave
analog--digital $L^2$ errors of about 155, 16 and 14~$\mu$V for unscaled,
single-range and two-range programming. Because the second bank added little,
the improved-buffer experiments below use a single range.

\section{Accuracy budget and physical operating range}\label{sec:budget}
\subsection{Circuit error relative to discretization error}
Let $u_h$ be the digital FEM field on the circuit mesh and $u_a$ the field
reconstructed from the decoded buffer outputs. We keep each element's own
outputs even in CG, because two buffers that sense the same port can have
different tracking errors. With the element mass matrix $M_e$, we define
\begin{equation}
 E_c^2=\norm{u_a-u_h}_{L^2}^2
 =\sum_e(u_{a,e}-u_{h,e})^TM_e(u_{a,e}-u_{h,e}),\qquad
 E_d=\norm{u_h-u_*}_{L^2},
\end{equation}
where $u_*$ is a reference field. We use the analytical solution where one is
available and a finer digital solution for elasticity. The figures write
$E_{\rm circ}$ and $E_{\rm disc}$ for $E_c$ and $E_d$, and call $100E_c/E_d$ the
budget ratio. On the unit cube the
volume RMS error equals the $L^2$ norm. For the bunny we divide both norms by
the square root of the volume to report RMS displacements, which leaves their
ratio unchanged. We require $E_c\le\eta E_d$ with $\eta=0.1$, so the circuit
may add at most one tenth of the error that the mesh already carries. Because
the denominator is the digital field's own error against the reference, a more
accurate mesh tightens the allowable circuit error.

\begin{proposition}[Total error under the circuit budget]
For $0\le\eta<1$, if $E_c\le\eta E_d$, then
\begin{equation}
 (1-\eta)E_d\le\norm{u_a-u_*}_{L^2}\le(1+\eta)E_d .
 \label{eq:paper-bound}
\end{equation}
If the same budget holds along a mesh sequence with $E_d\to0$, the
analog fields converge to $u_*$. If $E_d$ is of order $h^r$ with
positive upper and lower constants, the analog error has the same two-sided order.
\end{proposition}
\begin{proof}
Apply the triangle inequality and its reverse to
$u_a-u_*=(u_h-u_*)+(u_a-u_h)$.
\end{proof}
The bounds hold separately on each mesh. Their intervals can overlap, so two
successive refinements can both meet the budget even if the total error rises
slightly.

\subsection{Fixed-state memristor reads and voltage scaling}
For the coefficient branches we use the Stanford resistive random-access
memory (RRAM) read law at a fixed programmed gap
\citep{jiang2014rram,stanfordrramguide}. With zero-bias conductance $G$, the
current through a cell with terminal voltage difference $\Delta v$ is
\begin{equation}
 I(\Delta v)=G V_0\sinh(\Delta v/V_0),\qquad V_0=0.25~\mathrm{V}.
\end{equation}
The model defaults give a programmable range from about
$2.99~\mu\mathrm{S}$ to $1.80~\mathrm{mS}$. Each active branch keeps its
programmed state throughout the solve. The read cells load the buffer outputs,
while the signed-coefficient, face and current-decoding stages use ideal
sources.

To change the read voltage, we multiply the load currents and prescribed
voltages by $\alpha=A/A_{\rm ref}$, where $A_{\rm ref}=50~\mathrm{mV}$ in the
cube experiments, and recover each element's field as $u_{a,e}=D_e o_e/\alpha$
before computing errors. For a fixed reference-scale voltage difference
$\Delta\bar v$, the recovered branch current expands as
\begin{equation}
 \frac{I(\alpha\Delta\bar v)}{\alpha}
 =G\Delta\bar v+\frac{G\alpha^2(\Delta\bar v)^3}{6V_0^2}
 +O(\alpha^4).
 \label{eq:paper-read-expansion}
\end{equation}
Lowering the drive therefore reduces the leading read distortion
quadratically. Dividing the output voltage by $\alpha$, however, also magnifies
any absolute buffer tracking error that persists at low drive, while the
programmed conductances stay fixed. These competing effects motivate the drive
sweep in Section~\ref{sec:drive}. Throughout, we distinguish the nominal drive
$A$ from the voltage difference that actually appears across each read cell.

\subsection{Row buffer and stability tests}
Each row buffer is a voltage follower built from a cascoded NMOS differential
pair with a PMOS current mirror, complementary MOS level shifters and a
complementary common-source output stage. The simulations use GF180MCU 6~V
BSIM4 transistor models \citep{gf180models}, $\pm2.5$~V supply rails, ideal
bias references and a 33~pF Miller compensation capacitor. We selected the
design in isolated-buffer tests before evaluating any FEM field.

In 181 isolated-buffer AC tests, dual injection into the feedback loop
\citep{tianstability} gave minimum phase and gain margins of $52.8^\circ$ and
6.72~dB. We also tested all four P1/P2 CG and SIPG circuits on a
48-tetrahedron mesh by raising the body load from 99\% to 100\% of its nominal
value, starting from a biased operating point. In every case the row and
output voltages settled to within $1~\mu\mathrm{V}$ of an independent
full-load analog DC solution. These transients describe recovery on that small
mesh. The larger cube meshes are evaluated at their DC operating points.

\section{Controlled experiments}\label{sec:controlled}
\subsection{Digital references and circuit controls}
We compare each circuit with digital FEM on the same mesh and in the same
approximation space. CG shares unknowns at mesh nodes, whereas SIPG has a
separate set for every element. For each new P2 mesh and method we solve four
versions of the circuit in LTspice. The first uses ideal followers and linear
conductances that represent the coefficients exactly. The second imposes the
device conductance range, the third adds the nonlinear read law, and the
fourth replaces the ideal followers with transistor buffers. The $L^2$ distance
between successive fields measures the effect of each change, and these
effects can either reinforce or partly cancel one another.

An earlier uniform-lid benchmark, with linear coefficient resistors and the
original servo, provides a baseline. On the refined meshes of that benchmark,
the P1 and P2 CG circuits reproduce digital CG to \PoneCircuitMicroV{}
and \PtwoCircuitMicroV{}~$\mu$V volume RMS. The drive studies that follow use
the fixed-state read law and the improved servo. All four variants on a mesh
use the same requested coefficients and solver tolerances, and no circuit solve
starts from a digital FEM field. We checked the element matrices by independent
polynomial integration and verified current balance from the simulated branch
currents.

\subsection{Selecting a common read voltage}\label{sec:drive}
We first solve $-\Delta u=f$ on the unit cube with zero boundary values and the
exact solution $u_{*,50}=0.05\sin(\pi x)\sin(\pi y)\sin(\pi z)$ in volts, for
which $f=3\pi^2u_{*,50}$. We refine locally on three meshes whose largest
element diameter stays fixed. On each mesh we recompute the element and face
scales, vary only the source amplitude, and recover every field at the original
50~mV amplitude.

\begin{figure}[!htbp]\centering
\begin{subfigure}[t]{.485\linewidth}\centering
\includegraphics[width=\linewidth]{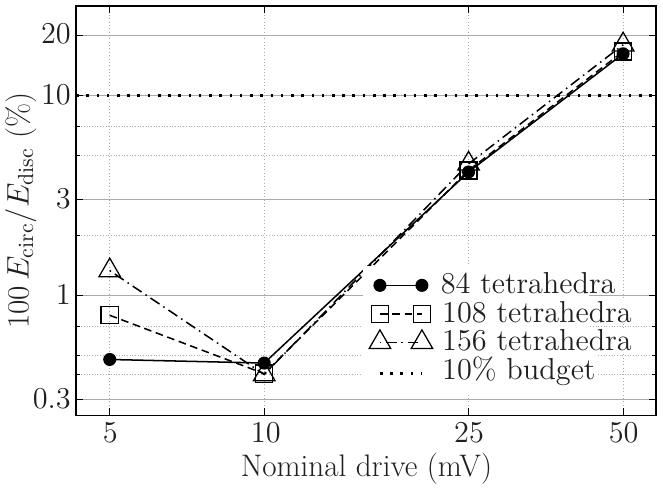}
\caption{P2 CG.}\label{fig:paper-voltage-cg}
\end{subfigure}\hfill
\begin{subfigure}[t]{.485\linewidth}\centering
\includegraphics[width=\linewidth]{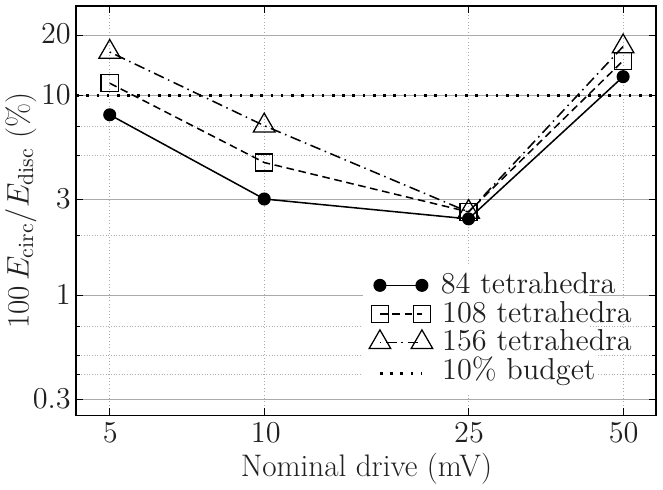}
\caption{P2 SIPG.}\label{fig:paper-voltage-sipg}
\end{subfigure}
\caption{Circuit error versus read voltage on three locally refined meshes.
The dotted line marks 10\% of the same-mesh digital discretization error.}
\label{fig:paper-voltage}
\end{figure}

We then select the largest tested drive at which both P2 methods meet the 10\%
budget on all three meshes (Figure~\ref{fig:paper-voltage}). All six cases pass
at 25 and 10~mV. At 50~mV all six exceed the budget, and at 5~mV the two finer
SIPG cases exceed it. On the finest SIPG mesh at 5~mV, replacing the transistor
buffers with ideal followers lowers the rescaled analog--digital error from 162
to 1.84~$\mu$V. Lowering the drive reduces read nonlinearity, but recovering
the physical voltage also magnifies the effect of buffer tracking error. We
carry 25~mV into the next test.

\subsection{Testing 25 mV with a new forcing and mesh family}
\begin{figure}[!htbp]\centering
\includegraphics[width=\linewidth]{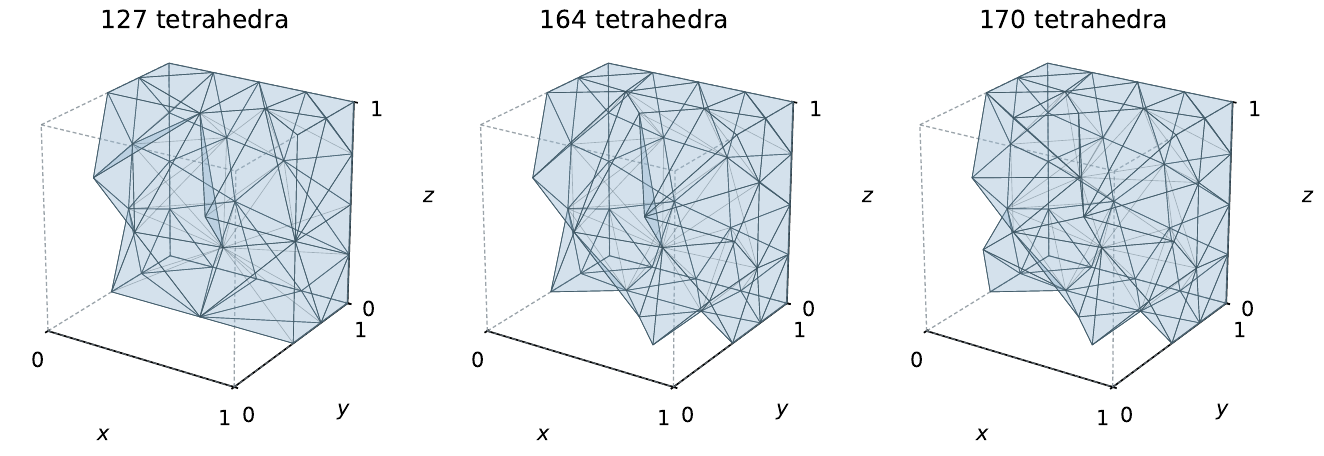}
\caption{Cutaway views of the three TetGen meshes. Only tetrahedra with
centroid $y\geq 1/2$ are shown. Both methods solve on the full cube.}
\label{fig:paper-new-meshes}
\end{figure}

Keeping the selected drive, we double the spatial frequency in $x$, so that the
new solution takes both positive and negative values. At the reference
amplitude it satisfies
\begin{equation}
 u_{*,50}=0.05\sin(2\pi x)\sin(\pi y)\sin(\pi z),\qquad
 f=6\pi^2u_{*,50},\qquad u_{*,50}|_{\partial\Omega}=0 .
 \label{eq:paper-new-solution}
\end{equation}
TetGen \citep{tetgenmanual} produced three distinct meshes with the same
asymmetric interior seed points (Figure~\ref{fig:paper-new-meshes}). We fixed
all three geometries before solving for any field, so they test the chosen
drive on geometries that played no part in selecting it. The element and face
gains follow the same geometry-dependent rules as before. We halve the
reference source amplitudes and decode the outputs at 50~mV.

\begin{figure}[!htbp]\centering
\includegraphics[width=.62\linewidth]{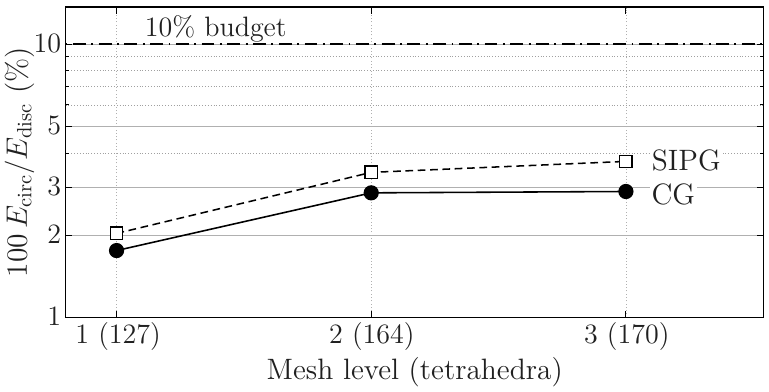}
\caption{P2 circuit error as a percentage of digital discretization error
on the three new meshes. The dash-dotted line marks the 10\% budget.}
\label{fig:paper-new-budget}
\end{figure}

At 25~mV, all six transistor-buffer circuits meet the budget. Their $L^2$
differences from digital FEM range from \GeneralizationMinRatioPercent{} to
\GeneralizationMaxRatioPercent\% of the matching discretization error
(Figure~\ref{fig:paper-new-budget}), and
Table~\ref{tab:p2-generalization-accuracy} also gives the total error against
the exact solution. The circuits with exact coefficients and ideal followers
reproduce the digital fields to numerical precision.

\begin{table}[!htbp]\centering\small
\caption{P2 field errors for the sign-changing solution at a 25~mV drive.}
\label{tab:p2-generalization-accuracy}
\begin{tabular}{r r r r r}\toprule
& \multicolumn{2}{c}{Circuit vs. digital ($\mu$V)} &
  \multicolumn{2}{c}{Circuit vs. exact (mV)}\\
\cmidrule(lr){2-3}\cmidrule(lr){4-5}
Tetrahedra & CG & SIPG & CG & SIPG\\\midrule
127 & 53.1 & 47.3 & 3.05 & 2.36\\
164 & 56.5 & 52.9 & 2.01 & 1.58\\
170 & 52.0 & 54.5 & 1.83 & 1.49\\
\bottomrule\end{tabular}
\end{table}

\begin{figure}[!htbp]\centering
\begin{subfigure}[t]{.485\linewidth}\centering
\includegraphics[width=\linewidth]{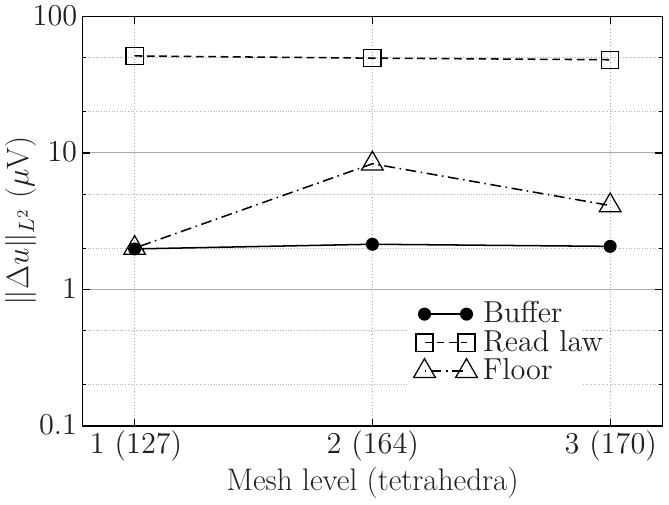}
\caption{P2 CG.}\label{fig:paper-controls-cg}
\end{subfigure}\hfill
\begin{subfigure}[t]{.485\linewidth}\centering
\includegraphics[width=\linewidth]{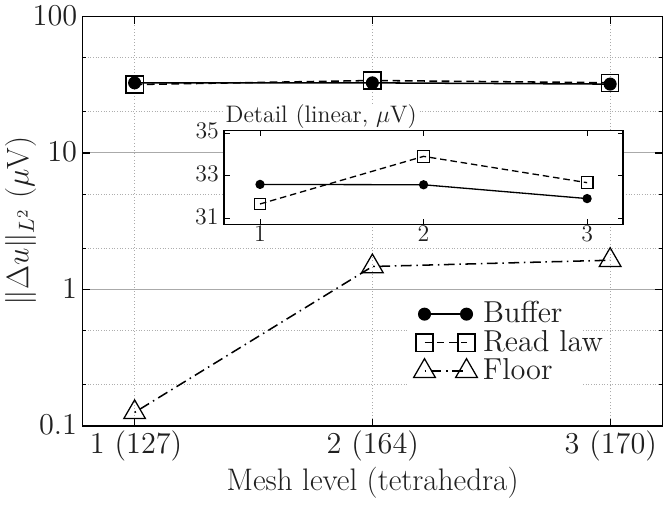}
\caption{P2 SIPG.}\label{fig:paper-controls-sipg}
\end{subfigure}
\caption{Pairwise $L^2$ field differences after imposing the conductance floor,
introducing the nonlinear read law, and replacing the ideal followers. The SIPG
inset shows the read and buffer curves on a linear scale.}
\label{fig:paper-controls}
\end{figure}

Figure~\ref{fig:paper-controls} shows which changes in the circuit matter most.
For CG, introducing the nonlinear read law changes the field far more than
replacing the ideal followers. For SIPG the two changes have similar sizes in
the full circuit, where the buffered outputs also control the face currents.
The conductance floor has the smallest effect in SIPG. Although the nominal
drive is 25~mV, individual read cells see voltage drops of up to 43~mV.

\FloatBarrier

\section{Refinement against common reference fields}\label{sec:refinement}
\input{paper_smooth_lid}
\input{paper_elasticity}

\section{What sets the operating range}\label{sec:operating}
The buffer count follows directly from the mesh and the polynomial order. A
scalar P2 tetrahedron has ten rows, so the 170-element cube uses 1,700 buffers
in either method, and SIPG also needs signed currents to couple
neighboring elements across faces. Component counts for every mesh are part of
the supplementary material. The simulations treat closed switches as ideal
connections and the signed gains and bias references as ideal circuit
elements, so the results describe nominal circuit operation. Estimating chip
area, power or settling time for a complete mesh would also require a physical
implementation of these components and their interconnect.

Two effects decide how the accuracy budget changes under refinement. A finer
mesh can bring the digital solution closer to the reference and so reduce the
allowable circuit error. At the same time, the mesh changes the conductances
and the loads that individual buffers see. The smooth-lid experiment separates
these effects by comparing both fields with an exact solution. The elasticity
experiment adds a second lesson. Even with exact coefficient resistors, the
voltage at a loaded buffer output can differ slightly from the port it senses,
and on finer meshes that small difference costs more of the budget. The signed
CG voltage encoding and the SIPG current correction address it within their
respective assemblies.

\section{Conclusion}\label{sec:conclusion}
A repeating tetrahedral circuit block can preserve Galerkin assembly while
letting each element use its own scale. In CG the current scaling is undone
before currents add at shared nodes, and in SIPG the change of coordinates is
applied consistently to the volume terms, the face terms and the loads. With
ideal coefficient conductances, the baseline P1 and P2 CG circuits reproduce
their digital fields to \PoneCircuitMicroV{} and \PtwoCircuitMicroV{}~$\mu$V
RMS.

With fixed-state memristor reads and transistor buffers, a 25~mV drive selected
on one mesh family met the 10\% budget on a second family with a different
forcing. On the smooth-lid cube, the total field error fell over three mesh
levels for both orders and both methods. In elasticity, reprogramming the P1 CG
elements brought the finest circuit within budget, and correcting the
difference between each row voltage and its buffer output in both the face and
the volume currents recovered the digital SIPG equilibrium with ideal
correction gains. Together these tests show where the repeating block meets a
mesh-dependent accuracy requirement, and how to find the circuit term to change
when refinement tightens that requirement.

\FloatBarrier
\section*{Data availability}
All circuits and data are available from the author upon reasonable request.
The supplementary notes and tables cited in the text are available on the same
basis.
\bibliographystyle{unsrtnat}
\bibliography{references}
\end{document}

%% file: generated/classical_values.tex
\newcommand{\PoneCircuitMicroV}{1.342}
\newcommand{\PtwoCircuitMicroV}{5.161}

%% file: generated/p2_generalization_values.tex
\newcommand{\GeneralizationMinRatioPercent}{1.76}
\newcommand{\GeneralizationMaxRatioPercent}{3.73}

%% file: paper_two_triangle.tex
\subsection{Two triangles sharing an edge}
\label{sec:two-triangle-example}
\begin{figure}[!htbp]
\centering
\begin{minipage}{.24\linewidth}\centering
\includegraphics[width=\linewidth,height=.132\textheight,keepaspectratio]{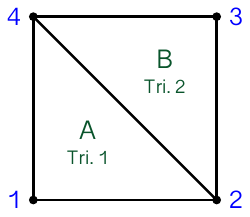}
\end{minipage}\hfill
\begin{minipage}{.72\linewidth}
\small\textbf{(a) Mesh and local stiffness matrices}\par\smallskip
\(A=(1,2,4)\), \(\quad B=(2,3,4)\)
\[
 K_A=\tfrac12\!\begin{bmatrix}2&-1&-1\\-1&1&0\\-1&0&1\end{bmatrix},
 \quad
 K_B=\tfrac12\!\begin{bmatrix}1&-1&0\\-1&2&-1\\0&-1&1\end{bmatrix}.
\]
\end{minipage}\par\medskip
{\raggedright\small\textbf{(b) Continuous Galerkin}\par}
\includegraphics[width=\linewidth]{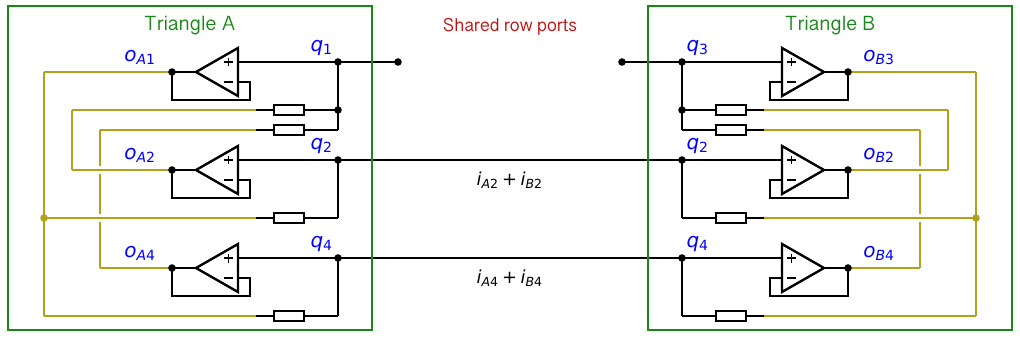}\par\smallskip
{\raggedright\small\textbf{(c) Discontinuous Galerkin}\par}
\includegraphics[width=\linewidth]{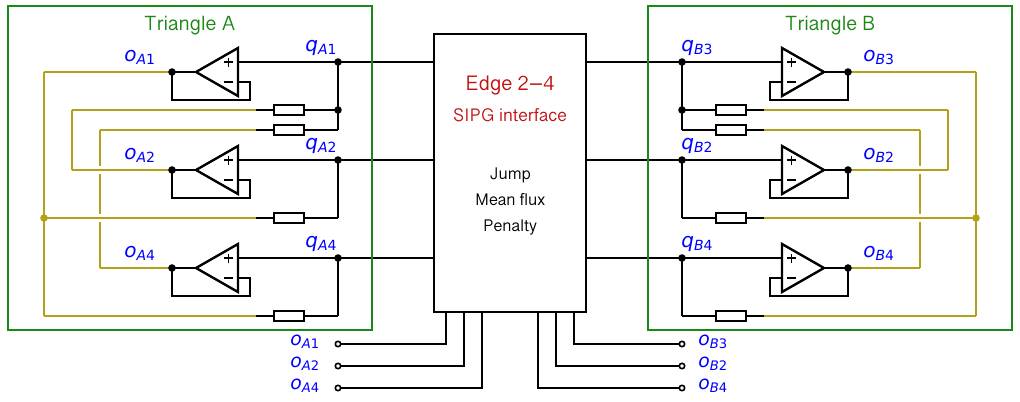}
\caption{Two P1 triangles sharing edge $2$--$4$.
CG joins matching row ports. SIPG couples the six separate rows through
the edge circuit. The diagrams omit exterior boundary connections and loads.}
\label{fig:two-triangle-example}
\end{figure}

Figure~\ref{fig:two-triangle-example} shows how the two assembly rules connect
a pair of P1 triangles. The diagonal from vertex~2 to vertex~4 divides a unit
square, and panel~(a) lists the two stiffness matrices for unit conductivity.
At a scale of $100~\mu\mathrm{S}$ per unit stiffness with $s_e=1$ and
$D_e=I$, each entry of $-1/2$ calls for a $20~\mathrm{k}\Omega$ resistor from
the receiving port $q_i$ to the sending output $o_j$. Each tile therefore uses
four such resistors and three servo buffers. Both tiles place the right-angle
vertex on the first pin, so the physical pin order is $(1,2,4)$ in A and
$(3,2,4)$ in B.

In CG the middle wire in panel~(b) joins A2 and B2 at the voltage $q_2$. With
negligible follower input current, the load current supplied at vertex~2 is
\[
 I_2=\frac{q_2-o_{A1}}{20\,\mathrm{k}\Omega}
     +\frac{q_2-o_{B3}}{20\,\mathrm{k}\Omega}.
\]
Both followers sense $q_2$, and each drives the coefficient branches of its own
tile. The same connection assembles vertex~4.

In SIPG the ports $q_{A2}$ and $q_{B2}$ stay separate. The interface in
panel~(c) senses outputs from both tiles and applies a current to each of the
six rows. With the normal from A to B and the jump $u_A-u_B$, the current
leaving row A1 through the edge circuit is
\[
 i^{\,2-4}_{A1}=50\,\mu\mathrm{S}\,(o_{A2}+o_{A4}-o_{B2}-o_{B4}).
\]
Although vertex~1 lies opposite the edge, its basis function has a nonzero
normal derivative there, so its row takes part in the coupling. The edge
penalty is $16/h_\perp$, where $h_\perp=1/\sqrt{2}$ is the altitude of either
triangle over the shared edge. The supplementary material traces the signed
currents on all six receiving rails of this interface and checks them against
an LTspice operating point.

\FloatBarrier

%% file: paper_smooth_lid.tex
\subsection{A smooth lid with an exact harmonic reference}
\label{sec:harmonic-lid}
A cube whose top face is driven uniformly has an abrupt change in prescribed
voltage where the top meets the grounded sides, and setting the edge nodes to
zero leaves that jump in the boundary data. To remove it, we keep the other
five faces grounded and prescribe $g(x,y)=V_0\sin(\pi x)\sin(\pi y)$ on the top
face, with $V_0=50$~mV. The voltage now peaks at the center and falls smoothly
to zero along all four edges. With no body source, separation of variables
gives the exact harmonic field
\begin{equation}
 u_*(x,y,z)=V_0\sin(\pi x)\sin(\pi y)
            \frac{\sinh(kz)}{\sinh k},\qquad k=\sqrt{2}\,\pi .
 \label{eq:harmonic-lid}
\end{equation}

The smooth and uniform lids share three grids of decreasing maximum diameter.
The first refinement adds a layer in depth, and the second refines both
horizontal directions. CG prescribes the voltage at boundary nodes, whereas
SIPG imposes it weakly through integrals over the boundary faces. Each pair of circuits uses
the same improved buffer, programmed conductances, and element and face scales,
so only the prescribed source values can differ. We drive both circuits at half
the reference amplitude and recover each local output in physical volts before
comparing fields.

\begin{figure}[!htbp]\centering
\begin{subfigure}[t]{.47\linewidth}\centering
\includegraphics[width=\linewidth,height=.183\textheight,keepaspectratio]{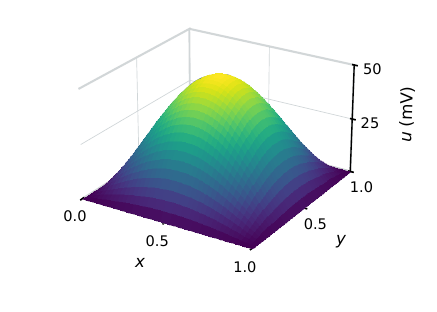}
\caption{Voltage on the lid.}
\end{subfigure}\hfill
\begin{subfigure}[t]{.47\linewidth}\centering
\includegraphics[width=\linewidth,height=.183\textheight,keepaspectratio]{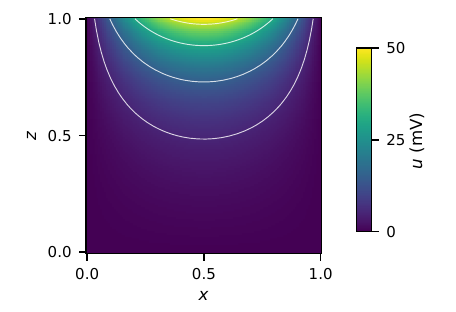}
\caption{Exact field at $y=1/2$.}
\end{subfigure}\par\medskip
\begin{subfigure}[t]{.97\linewidth}\centering
\includegraphics[width=\linewidth]{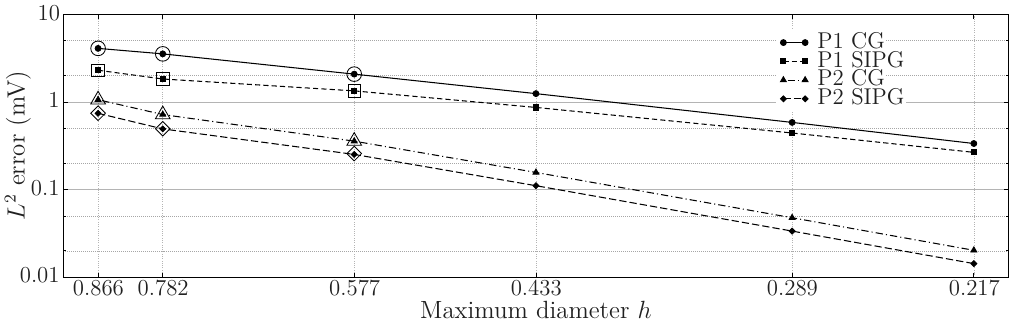}
\caption{Error against the exact smooth-lid field.}
\end{subfigure}\par\medskip
\begin{subfigure}[t]{.485\linewidth}\centering
\includegraphics[width=\linewidth]{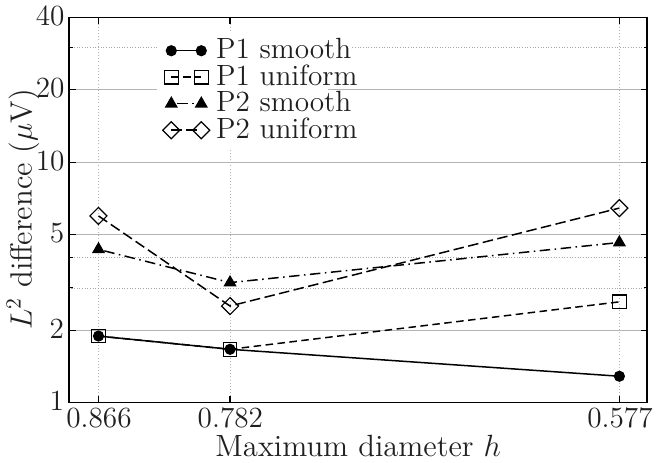}
\caption{Circuit vs. digital CG.}
\end{subfigure}\hfill
\begin{subfigure}[t]{.485\linewidth}\centering
\includegraphics[width=\linewidth]{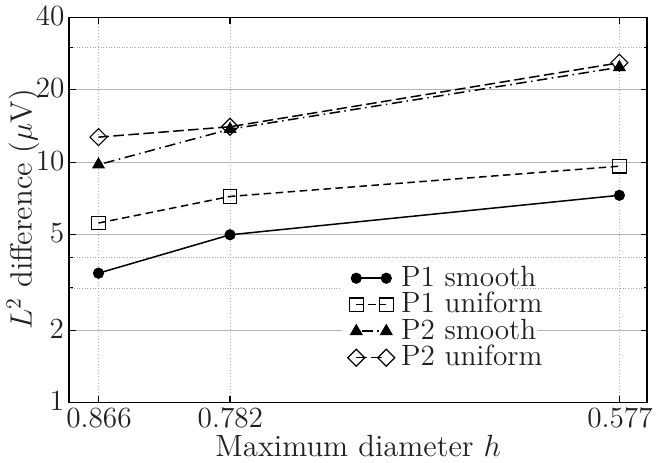}
\caption{Circuit vs. digital SIPG.}
\end{subfigure}
\caption{Harmonic cube solution and refinement errors. In the three line plots,
mesh diameter decreases to the right.}
\label{fig:harmonic-lid}
\end{figure}

With the smooth lid, the total error falls at both refinements for all four
methods, and every circuit meets its same-mesh error budget.
Figure~\ref{fig:harmonic-lid}(c) follows the digital solution on three further
grids, with open symbols for the LTspice results on the first three, and
Table~\ref{tab:harmonic-lid} lists the circuit errors on the coarsest and
finest circuit meshes. P2 SIPG has the lowest total error on every circuit
mesh. Its digital discretization error is also the smallest, so on the finest
grid it comes closest to the 10\% budget.

\begin{table}[!htbp]\centering\small
\caption{Smooth-lid circuit errors on the coarsest and finest meshes.
The last two columns compare the finest circuit with same-mesh digital FEM.}
\label{tab:harmonic-lid}
\begin{tabular}{l r r r r}\toprule
& \multicolumn{2}{c}{Circuit vs. exact (mV)} &
  \multicolumn{2}{c}{Finest mesh}\\
\cmidrule(lr){2-3}\cmidrule(lr){4-5}
Method & 48 tetrahedra & 162 tetrahedra & $E_c$ ($\mu$V) & $100E_c/E_d$ (\%)\\\midrule
P1 CG & 4.089 & 2.078 & 1.29 & 0.06\\
P2 CG & 1.062 & 0.358 & 4.63 & 1.30\\
P1 SIPG & 2.301 & 1.340 & 7.28 & 0.54\\
P2 SIPG & 0.744 & 0.258 & 24.82 & 9.86\\
\bottomrule\end{tabular}
\end{table}

The difference between P2 SIPG and same-mesh digital FEM grows with refinement
for both lids (Figure~\ref{fig:harmonic-lid}(e)). To find out why, we convert
the measured face-current error $\diag(F)(q-o)$ into physical load units and
solve the digital SIPG system for its contribution to the field. On the smooth
lid this contribution grows from about 8 to 21~$\mu$V, even though the mean
tracking error $q-o$ stays near $-0.10~\mu\mathrm{V}$, and the uniform lid
gives nearly the same range. This face term is the largest contributor on all
three meshes, and it returns with greater force in the elasticity study below.

%% file: paper_elasticity.tex
\subsection{Elastic displacement under an ear load}
\label{sec:bunny-refinement}
Our second application is three-dimensional linear elasticity. We model the
Stanford bunny \citep{stanfordbunny} as an isotropic elastic solid $\Omega$,
0.1~m tall, with Young's modulus $E=1$~MPa and Poisson's ratio $\nu=0.3$. On
the initial mesh, the exterior faces whose centroids lie within 5~mm of the
base plane form the clamped boundary $\Gamma_D$. A force $\mathbf F$ of 1~mN
acts in the positive $x$ direction at the highest boundary vertex
$\mathbf x_0$ on the ear, and the rest of the surface, $\Gamma_N$, is traction
free. Refinement preserves the clamped faces and the loaded vertex.

\paragraph{Governing equations.}
The displacement $\mathbf u$ satisfies the equations of static linear
elasticity,
\begin{align}
 -\nabla\cdot\boldsymbol\sigma(\mathbf u)&=\mathbf 0
 \quad\text{in }\Omega, \label{eq:el-equilibrium}\\
 \boldsymbol\sigma(\mathbf u)&=\lambda(\nabla\cdot\mathbf u)\,\mathbf I
 +2\mu\,\boldsymbol\varepsilon(\mathbf u),\qquad
 \boldsymbol\varepsilon(\mathbf u)=\tfrac12\bigl(\nabla\mathbf u
 +\nabla\mathbf u^{T}\bigr), \label{eq:el-constitutive}\\
 \mathbf u&=\mathbf 0\quad\text{on }\Gamma_D,\qquad
 \boldsymbol\sigma(\mathbf u)\mathbf n=\mathbf 0\quad\text{on }\Gamma_N,
 \label{eq:el-bc}
\end{align}
where $\mathbf n$ is the outward unit normal and the Lam\'e parameters are
\begin{equation}
 \lambda=\frac{E\nu}{(1+\nu)(1-2\nu)}\approx0.577~\mathrm{MPa},\qquad
 \mu=\frac{E}{2(1+\nu)}\approx0.385~\mathrm{MPa}.
 \label{eq:el-lame}
\end{equation}
Substituting the constitutive law into the equilibrium equation gives the
Navier form $(\lambda+\mu)\nabla(\nabla\cdot\mathbf u)+\mu\Delta\mathbf
u=\mathbf 0$. A concentrated force has no finite-energy continuum solution, so
we define the ear load at the mesh level through the work it does on a
discrete displacement field, $\ell(\mathbf v)=\mathbf F\cdot\mathbf
v(\mathbf x_0)$. In a control case the same force acts instead as the uniform
traction $\mathbf F/|\Gamma_P|$ on the patch $\Gamma_P$ of boundary faces that
meet at $\mathbf x_0$ on the initial mesh. This patch stays fixed during
refinement.

\paragraph{Continuous Galerkin form.}
Let $V_h$ contain the continuous vector fields that are linear on every
tetrahedron and vanish on $\Gamma_D$. The CG solution $\mathbf u_h\in V_h$
satisfies
\begin{equation}
\begin{aligned}
 a(\mathbf u_h,\mathbf v)&=\mathbf F\cdot\mathbf v(\mathbf x_0)
 \qquad\text{for all }\mathbf v\in V_h,\\
 a(\mathbf u,\mathbf v)&=\int_\Omega\boldsymbol\sigma(\mathbf u):
 \boldsymbol\varepsilon(\mathbf v)\,dV
 =\int_\Omega\bigl[\lambda(\nabla\cdot\mathbf u)(\nabla\cdot\mathbf v)
 +2\mu\,\boldsymbol\varepsilon(\mathbf u):\boldsymbol\varepsilon(\mathbf v)
 \bigr]dV .
\end{aligned}
 \label{eq:el-cg}
\end{equation}
Each node $a$ carries three vector basis functions $N_a\mathbf e_i$, one for
each displacement component $i$. Their element stiffness entries are
\begin{equation}
 (K_e)_{ai,bj}=\int_{T_e}\Bigl[\lambda\,\partial_iN_a\,\partial_jN_b
 +\mu\bigl(\delta_{ij}\,\nabla N_a\cdot\nabla N_b
 +\partial_jN_a\,\partial_iN_b\bigr)\Bigr]dV ,
 \label{eq:el-element}
\end{equation}
which form a symmetric $12\times12$ matrix on a P1 tetrahedron, where the
gradients are constant, and a $30\times30$ matrix on a P2 tetrahedron. Rigid
translations and rotations produce no strain, so $K_e$ has a six-dimensional
null space, the elastic counterpart of $K_e\bm1=0$. In the circuit, each P1
element block has twelve rows, one for each displacement component at each
vertex. Joining the row ports of each displacement component at every shared
vertex assembles $K=\sum_eR_e^TK_eR_e$ exactly as in~\eqref{eq:paper-assembly}, now with three
rows per mesh node and with the element gains decoded as
in~\eqref{eq:paper-decoding}. The clamp is imposed strongly by holding the
rows of every node on $\Gamma_D$ at zero volts, and the ear force enters as a
current injected at the $x$ row of the loaded vertex.

\paragraph{SIPG form.}
In SIPG the displacement is linear on each tetrahedron but may jump across
faces, and each element keeps its own twelve rows. On an interior face we use
the common normal $\mathbf n$ from the left to the right element, the vector
jump $[\mathbf u]=\mathbf u_L-\mathbf u_R$ and the average traction
$\{\boldsymbol\sigma(\mathbf u)\mathbf n\}=\tfrac12\bigl(\boldsymbol\sigma
(\mathbf u_L)+\boldsymbol\sigma(\mathbf u_R)\bigr)\mathbf n$. On a clamped face
we take $[\mathbf u]=\mathbf u$ and the one-sided traction with the outward
normal. With $\mathcal F_I$ the interior faces and $\mathcal F_D$ the clamped
faces, the discrete solution satisfies
$a_{\rm DG}(\mathbf u_h,\mathbf v)=\ell_{\rm DG}(\mathbf v)$ for every
piecewise linear $\mathbf v$, where
\begin{equation}
\begin{split}
 a_{\rm DG}(\mathbf u,\mathbf v)={}&\sum_e\int_{T_e}\boldsymbol\sigma(\mathbf u):
 \boldsymbol\varepsilon(\mathbf v)\,dV\\
 &+\sum_{F\in\mathcal F_I\cup\mathcal F_D}\int_F\Bigl(
 -\{\boldsymbol\sigma(\mathbf u)\mathbf n\}\cdot[\mathbf v]
 -\{\boldsymbol\sigma(\mathbf v)\mathbf n\}\cdot[\mathbf u]
 +\tau_F^{\rm el}\,[\mathbf u]\cdot[\mathbf v]\Bigr)dS
\end{split}
 \label{eq:el-sipg}
\end{equation}
and
\begin{equation}
 \tau_F^{\rm el}=\frac{4(p+1)^2(\lambda+2\mu)}{\min(h_{L,F},h_{R,F})},\qquad
 \ell_{\rm DG}(\mathbf v)=\frac{1}{n_0}\sum_{e\ni\mathbf x_0}
 \mathbf F\cdot\mathbf v_e(\mathbf x_0).
 \label{eq:el-penalty-load}
\end{equation}
The penalty is the scalar penalty of~\eqref{eq:paper-sipg-face} multiplied by
the P-wave modulus $\lambda+2\mu$, with $p=1$ and the single available altitude
on a clamped face. Traction-free faces contribute nothing, and because the
clamped displacement is zero, the Nitsche terms on $\Gamma_D$ add no load. The
load functional splits the ear force equally among the $n_0$ element traces
that meet at $\mathbf x_0$. In the circuit, the volume terms are programmed
inside each element block, and signed face currents couple the twelve rows of
each element to its own outputs and to those of its neighbors, as in
Section~\ref{sec:dg-faces} with the scalar jump and normal derivative replaced
by the vector jump and traction. The row scales $D_e$ and element gains $s_e$
follow~\eqref{eq:paper-dg-scales} for each $12\times12$ element matrix.

\paragraph{Circuit encoding and solution.}
Displacements are encoded as voltages $q=\alpha_u u$ in CG and
$q_e=\alpha_uD_e^{-1}u_e$ in SIPG, with $\alpha_u=25{,}000$~V/m, so one
micrometre of CG displacement corresponds to 25~mV. Stiffness is
normalized by $K_{\rm ref}=EL=10^5$~N/m with $L=0.1$~m and converted to
conductance with the same factor $\gamma\approx1.56$~mS as in the diffusion
circuits. A
further factor $\kappa=0.1$ scales every coefficient conductance and the load
current alike. It leaves the ideal equilibrium unchanged but reduces the current
each buffer must supply. In CG the 1~mN ear load therefore enters as a current
$\kappa\gamma\alpha_uF/K_{\rm ref}\approx39$~nA at the loaded vertex. In SIPG
the same load is split among the element traces and multiplied by their row
factors, as in~\eqref{eq:paper-dg-transform}. The circuits use exact
coefficient resistors and the improved GF180MCU servo, and we update the volume
and face coefficients with each mesh. Because a fully coupled solve of the
larger networks can exceed the simulator's resources, a sparse outer solver
enforces current balance across the wired network. Each update evaluates the
nonlinear transistor response of every buffer in native LTspice at that row's
actual input voltage and output load. On the coarsest mesh we also checked the
SIPG field against a single LTspice solve of the complete circuit.

\paragraph{Error measure.}
The point load makes the displacement near the ear tip sensitive to the mesh,
so we compare displacement vectors throughout the solid. For each surface and
load, the reference solves~\eqref{eq:el-cg} digitally with continuous quadratic
(P2) fields on a finer nested mesh, with the same clamp and load. On every tetrahedron of that fine
mesh we evaluate the reference and the field of its parent element, keeping the
separate SIPG traces, and integrate the squared vector difference with the
consistent mass matrix. The denominator of the circuit budget is therefore a
volume-integrated discretization error rather than a value at the tip. Most of
it still comes from the loaded ear. About 90\% of its square lies within 20~mm
of $\mathbf x_0$, a region that holds less than 1\% of the volume.

\paragraph{First three meshes.}
We first hold the 320-facet surface, the base and the ear vertex fixed.
In panels (a) and (b) of Figure~\ref{fig:bunny-reference}, the circuit and
affine-model results cover the first three mesh levels. Both digital methods
approach the reference as the mesh is refined. The CG circuit
meets the 10\% budget at these levels, but the SIPG circuit exceeds it at all
three. With an affine servo model fitted to the DC measurements, replacing the
point load by the traction patch still leaves a substantial SIPG error.

The circuit equations explain the difference. The diagonal face coefficient
senses the row input $q_i$, whereas the off-diagonal coefficients sense buffer
outputs, so a tracking difference $q_i-o_i$ adds the row current
$F_{ii}(q_i-o_i)$. On the coarsest mesh the median tracking difference is only
$-104$~nV. Yet when we convert the extra currents into forces and solve for the
displacements they cause, the six most compliant elastic modes account for over
99\% of the squared total displacement error. Small voltage errors can
therefore drive appreciable motion of the ear and the body.

\subsection{Mesh refinement and surface detail}
\label{sec:bunny-fine}
We extend the point-load comparison to five P1 circuit meshes on the same
surface. The largest has 60,213 tetrahedra and 722,556 local buffers per
method. The new P2 reference has 181,869 tetrahedra. Its displacement differs
from the preceding P2 field by 1.04\% in the volume $L^2$ norm, and either
reference leads to the same budget decisions.

\begin{figure}[!htbp]\centering
\begin{subfigure}[t]{.485\linewidth}\centering
\includegraphics[width=\linewidth]{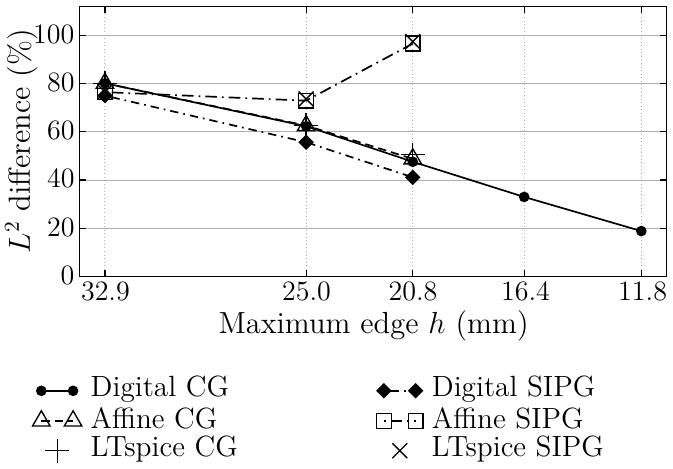}
\caption{Point load against P2 on 60,213 tetrahedra.}\label{fig:bunny-reference-common}
\end{subfigure}\hfill
\begin{subfigure}[t]{.485\linewidth}\centering
\includegraphics[width=\linewidth,trim=0 -2bp 0 0,clip]{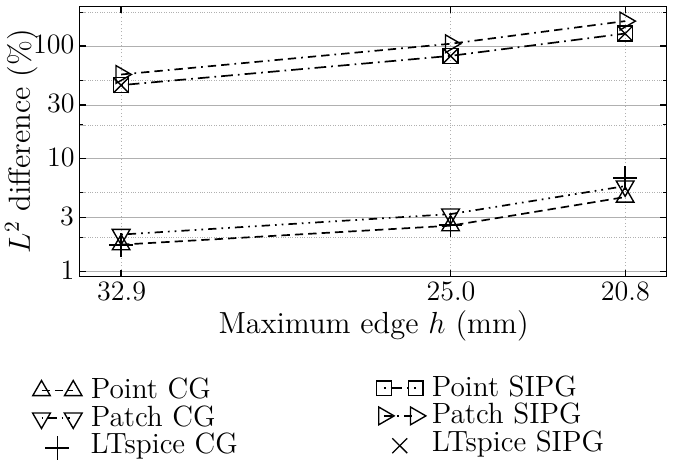}
\caption{Point and patch loads against same-mesh P1.}\label{fig:bunny-reference-same}
\end{subfigure}\par\medskip
\begin{subfigure}[t]{.485\linewidth}\centering
\includegraphics[width=\linewidth,trim=-5bp 0 0 0,clip]{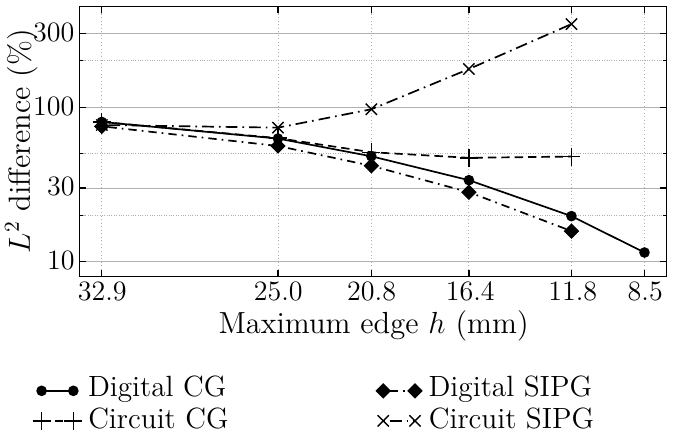}
\caption{Point load against P2 on 181,869 tetrahedra.}\label{fig:bunny-fine-common}
\end{subfigure}\hfill
\begin{subfigure}[t]{.485\linewidth}\centering
\includegraphics[width=\linewidth,trim=0 -2bp 0 0,clip]{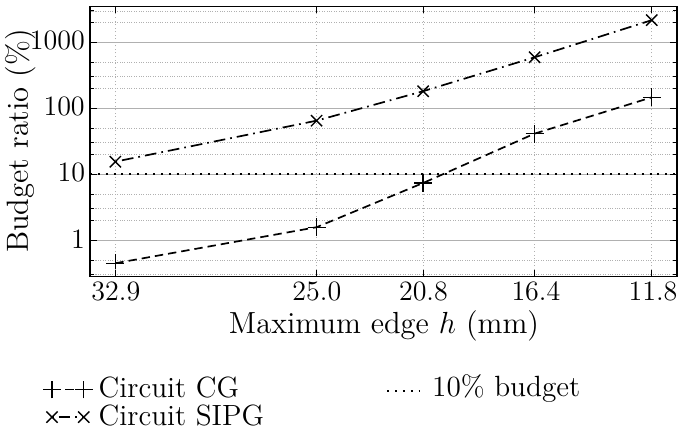}
\caption{Circuit error relative to discretization error.}\label{fig:bunny-fine-budget}
\end{subfigure}
\caption{P1 displacement errors under refinement of the fixed bunny surface.
Panels (a) and (c) use the indicated P2 reference. In (b) the affine servo
model covers point and patch loads, and LTspice symbols denote the point
load. Panel (d) gives circuit error relative to estimated discretization error.}
\label{fig:bunny-reference}
\end{figure}

On the finest circuit mesh the digital P1 errors are about 20\% for CG and 16\%
for SIPG relative to the finer P2 field. At this refinement the baseline
circuits exceed the budget (panels (c) and (d) of
Figure~\ref{fig:bunny-reference}). CG meets it on the first three meshes and
exceeds it on the last two, while SIPG exceeds it on all five. In CG the most
heavily loaded buffers cluster near the ear, and in SIPG the extra face current
continues to dominate the difference from digital P1. These observations
motivate the two changes tested in Section~\ref{sec:bunny-compensation}.

\begin{figure}[!htbp]\centering
\includegraphics[width=.98\linewidth]{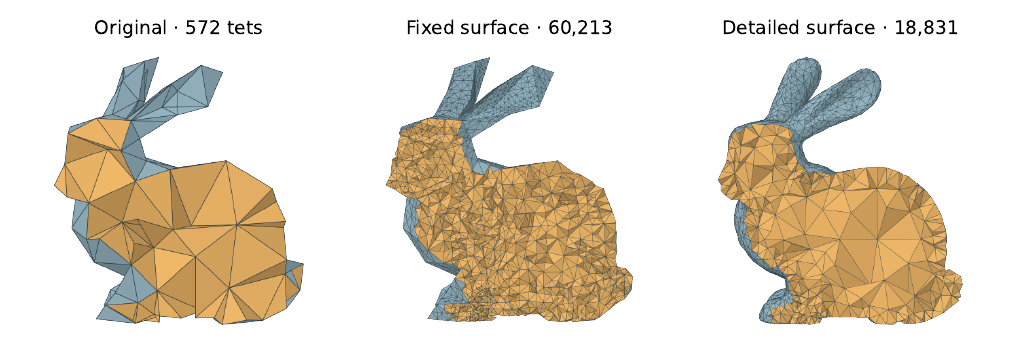}\par\smallskip
\includegraphics[width=.98\linewidth]{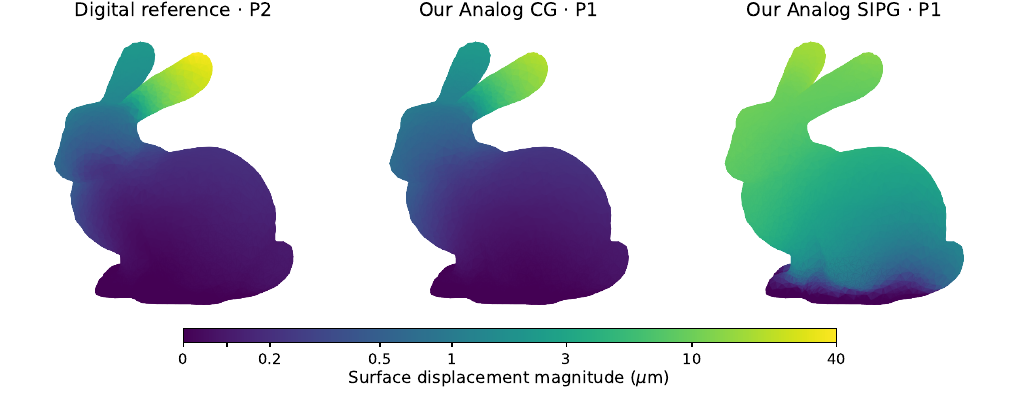}
\caption{Volume meshes on the original and detailed bunny surfaces (top)
and baseline displacements on the detailed surface (bottom).
Blue-gray marks exterior facets and ochre marks exposed interior facets.
Color in the lower panels gives displacement magnitude.}
\label{fig:bunny-fine-fields}
\end{figure}

We also tetrahedralize a more detailed, 6,000-facet version of the scan, select
the clamp and ear load by the same rules, and compute a new digital P2
reference on this geometry. Figure~\ref{fig:bunny-fine-fields} compares the
volume meshes and shows the baseline displacement fields on the detailed
surface. Each surface triangle takes its color from the displacement at its
centroid on the incident tetrahedron, which preserves the SIPG trace on that
element.

\subsection{Reducing buffer tracking error in elasticity}
\label{sec:bunny-compensation}
We test two changes with the same buffers, exact coefficient resistors, loads
and digital references. For CG we change how displacements are encoded and how
strongly each element loads its buffers. For SIPG we add a local current that
compensates for the measured difference between a row input and its buffer
output.

\begin{figure}[!htbp]\centering
\begin{subfigure}[t]{.485\linewidth}\centering
\includegraphics[width=\linewidth]{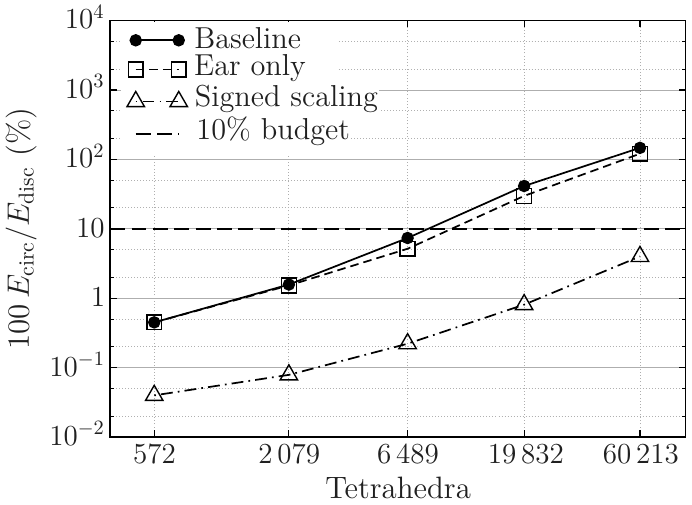}
\caption{CG scaling.}\label{fig:bunny-compensation-cg}
\end{subfigure}\hfill
\begin{subfigure}[t]{.485\linewidth}\centering
\includegraphics[width=\linewidth]{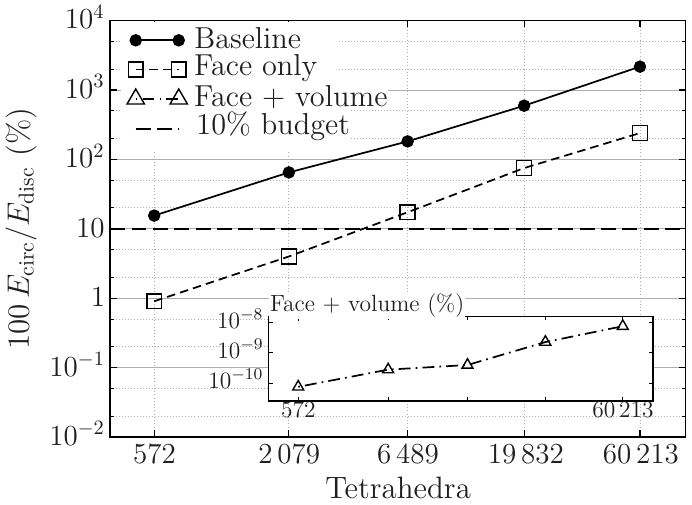}
\caption{SIPG tracking correction.}\label{fig:bunny-compensation-sipg}
\end{subfigure}\par\smallskip
\includegraphics[width=.97\linewidth]{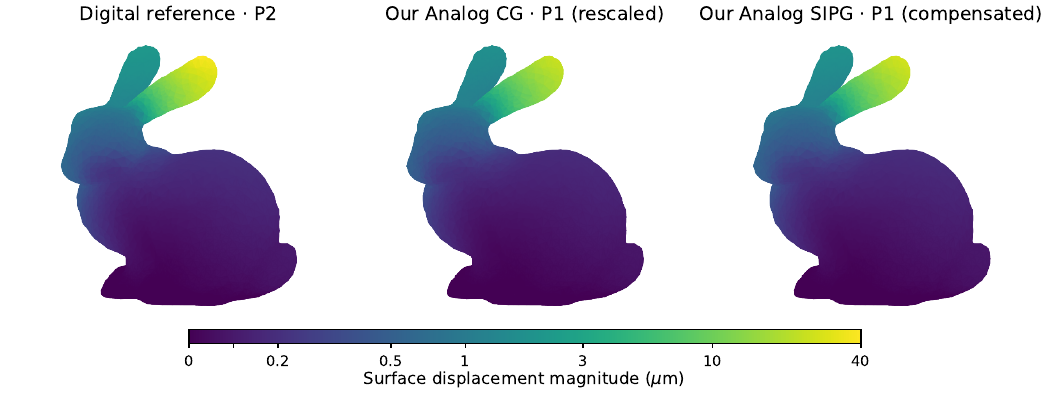}
\caption{CG scaling and SIPG correction on five fixed-domain meshes (top).
The inset resolves the combined SIPG correction. On the detailed surface
(bottom), color gives displacement for the P2 reference and the two
corrected P1 circuits.}
\label{fig:bunny-compensation}
\end{figure}

\paragraph{CG voltage and current scaling.}
We use the previous analog solution for the same mesh and ear load to identify
heavily loaded buffers. As a control, we halve the coefficient currents only
in elements whose centroid lies within 20~mm of $\mathbf x_0$ and whose largest
previous output load exceeded 100~$\mu$A, and double their decoding factors
$1/s_e$ at assembly. The full
method acts on the whole mesh. If the largest predicted output load in
element $e$ exceeds $10~\mu$A, we scale its coefficient currents by
$a_e=10~\mu\mathrm A/\max_j|I^{\rm est}_{ej}|$, and otherwise $a_e=1$.
Dividing each outgoing row current by $a_e$ preserves the element's weight in
the assembled system. Because the threshold is applied to the previous field,
the load at the new equilibrium can differ.

The full method also gives each shared displacement component a signed
voltage scale. We encode $u_i$ as $q_i=\alpha_um_iu_i$ and
choose $|m_i|$ so that the previous field would encode at or below 0.14~V,
leaving $|m_i|=1$ for smaller signals. To reduce the effect of a common buffer
offset, the programming algorithm balances the signs of neighboring coefficient
contributions for each displacement component, and every element that shares a
component uses the same $m_i$. With the diagonal matrix $P_e$ of these scales,
the programmed stiffness and load are $P_e^{-1}K_eP_e^{-1}$ and $P_e^{-1}b_e$,
and dividing each output by $\alpha_um_i$ recovers the displacement in metres.
Each new circuit solve starts from zero.

\paragraph{SIPG correction currents.}
Let $c_i^F$ be the signed diagonal face coefficient of row $i$ in siemens after
scaling. The face circuit draws the additional current $c_i^FH(s)(o_i-q_i)$
from row $i$, where $H$ is a low-pass filter with a 100~ns time constant and
unit DC gain. At equilibrium this cancels the extra face current. The
coefficient resistors also see the row input, because a resistor with
conductance $G_{ij}$ draws $G_{ij}(q_i-o_j)$, which contains the tracking term
$G_{ij}(q_i-o_i)$. Replacing $c_i^F$ by $c_i^F+\sum_jG_{ij}$ in the correction
cancels both terms. With ideal differential sensing and exact signed gains, the
DC equations for the buffer outputs then coincide with the digital SIPG
equations. Some corrected cases at 25,000~V/m exceeded the 0.5~V limit at the
buffer terminals, so we repeated all six corrected SIPG cases at 17,500~V/m,
with the load current and the decoding scaled to match. Every terminal then
stayed within the limit.

\begin{table}[!htbp]\centering\small
\caption{Displacement errors after CG scaling and SIPG correction.
The last column gives the circuit error as a percentage of the digital
error against the finer P2 reference.}\label{tab:bunny-compensated}
\begin{tabular}{lr lrrr}\toprule
Surface & Tetrahedra & Method & \multicolumn{2}{c}{Volume RMS error ($\mu$m)} & Ratio (\%)\\
\cmidrule(lr){4-5}
 & & & Circuit vs.\ P1 & P1 vs.\ P2 & \\ \midrule
Original & 60,213 & CG   & 0.0177 & 0.445 & 3.98 \\
         &        & SIPG & $<10^{-9}$ & 0.356 & $<10^{-4}$ \\
Detailed & 18,831 & CG   & 0.00749 & 0.586 & 1.28 \\
         &        & SIPG & $<10^{-9}$ & 0.477 & $<10^{-4}$ \\
\bottomrule\end{tabular}
\end{table}

Figure~\ref{fig:bunny-compensation} compares these choices. Scaling only the heavily loaded elements near the
ear brings a modest improvement on the finest original mesh. Balancing the
voltage signs as well as scaling the currents brings CG within budget on all six
meshes tested. In SIPG the face correction alone exceeds the budget on the finer
meshes, whereas correcting both the face and the volume currents recovers the
matching digital field to solver precision (Table~\ref{tab:bunny-compensated}).
This agreement tests the algebraic cancellation under ideal sensing. On the
finest mesh, an offset of only 0.36~nV shared by all equally oriented
correction sensors would exhaust the budget.

\paragraph{Circuit checks.}
For the full-mesh DC solutions we checked current balance from the circuit
components and recomputed the SIPG jumps and tractions from the mesh. On a pair
of tetrahedra near the ear, the signed CG circuit returned to its local
tolerance band within 0.75~ms after a load pulse and a current perturbation,
and the corrected SIPG circuit within 1.72~ms. These times describe the
two-element patch only, and the full-mesh comparisons measure DC equilibrium.

\subsection{Interior displacement, stress and ear deflection}
\label{sec:bunny-interior}
\begin{figure}[!htbp]\centering
\includegraphics[width=.90\linewidth]{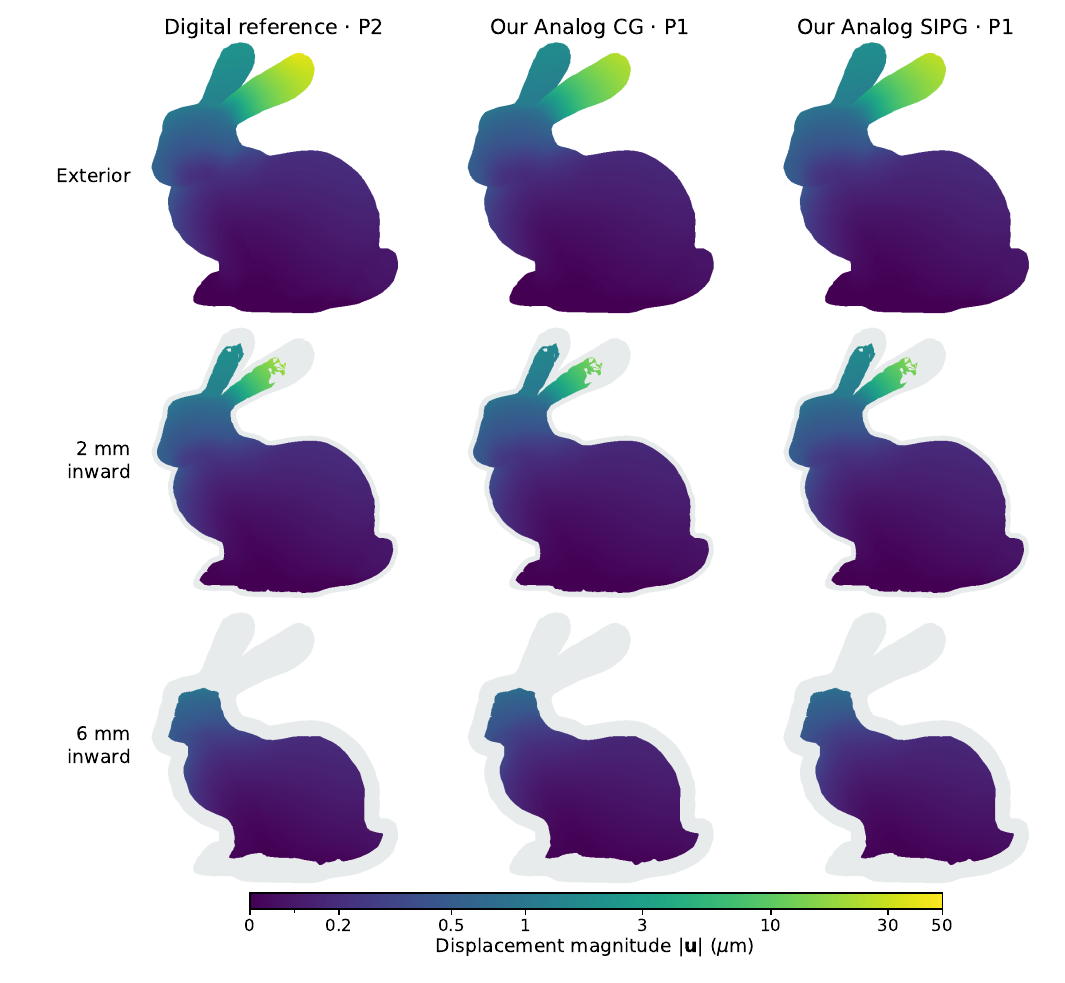}
\caption{Displacement magnitude on the exterior and at depths of 2 and
6~mm. All panels use the same color scale. Gray shows the original outline.}
\label{fig:bunny-interior-u}
\end{figure}

Figure~\ref{fig:bunny-interior-u} shows how far the displacement at the loaded
ear extends into the body. We compare the digital P2 reference on 73,296
tetrahedra with the scaled CG and corrected SIPG circuits on 18,831
tetrahedra, all on the same detailed surface. The interior views cut each field
at a constant distance below the exterior surface, so the cuts expose the same
anatomy in all three fields. The thin ears disappear from the deeper cut, and
the largest displacement visible in the remaining body and neck is about
$1~\mu$m in each solution.

\begin{figure}[!htbp]\centering
\includegraphics[width=.90\linewidth]{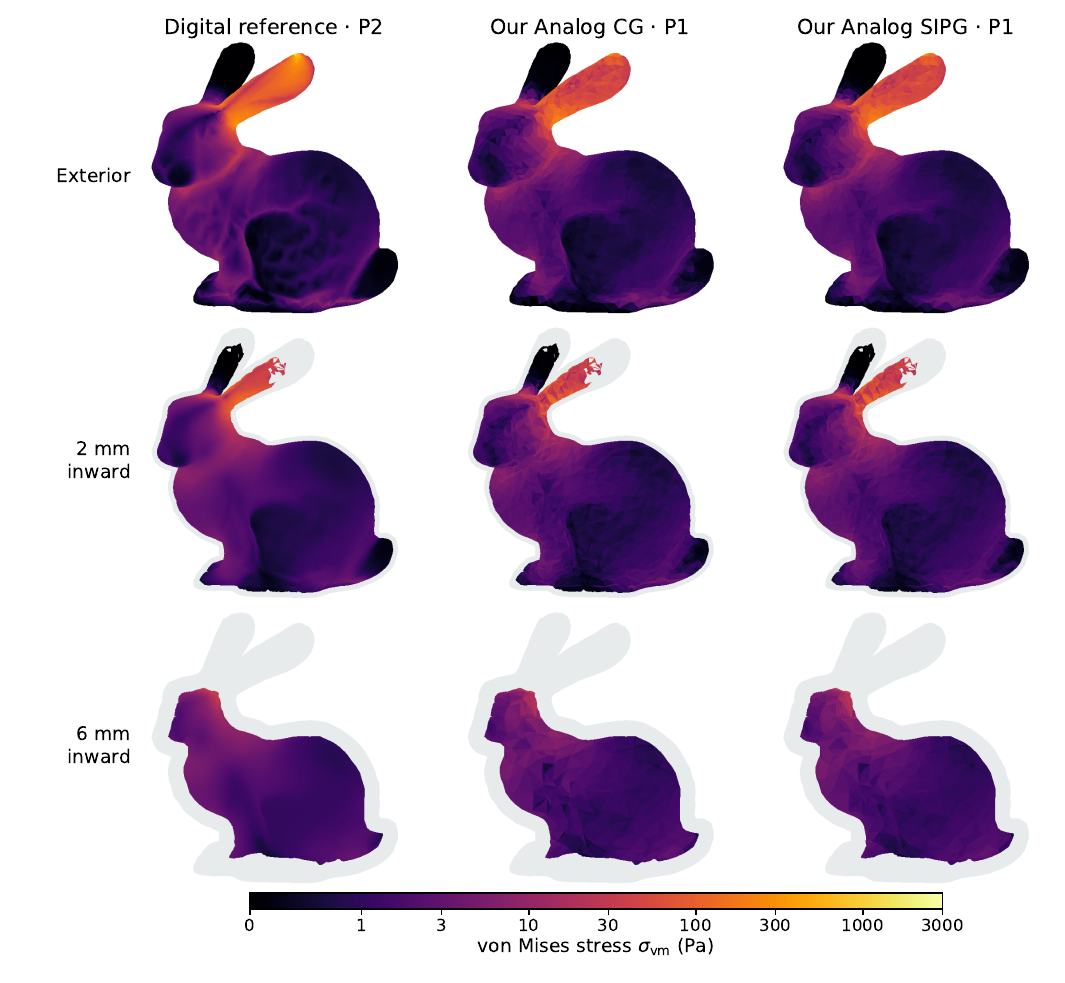}
\caption{Elementwise von Mises stress on the surfaces of
Figure~\ref{fig:bunny-interior-u}. All nine views use the same scale in pascals.}
\label{fig:bunny-interior-vm}
\end{figure}

We compute stress from the physical displacement on the original geometry
with~\eqref{eq:el-constitutive}, and summarize it by the von Mises stress
\begin{equation}
 \sigma_{\rm vM}=\sqrt{\tfrac32\,\mathbf s:\mathbf s},\qquad
 \mathbf s=\boldsymbol\sigma-\tfrac13\operatorname{tr}(\boldsymbol\sigma)\,\mathbf I ,
 \label{eq:el-vonmises}
\end{equation}
evaluated within each element (Figure~\ref{fig:bunny-interior-vm}). The stress
tensor is constant on a P1 tetrahedron and linear on a P2 tetrahedron, and the
local stress peaks near the concentrated load depend on mesh resolution. Over
the whole volume, the RMS difference in von Mises stress between analog CG and
same-mesh digital CG is 0.43\% of the digital RMS von Mises stress. Under the ideal sensing
assumption, the compensated SIPG stress agrees with its digital counterpart to
solver precision.

\begin{figure}[!htbp]\centering
\includegraphics[width=.90\linewidth]{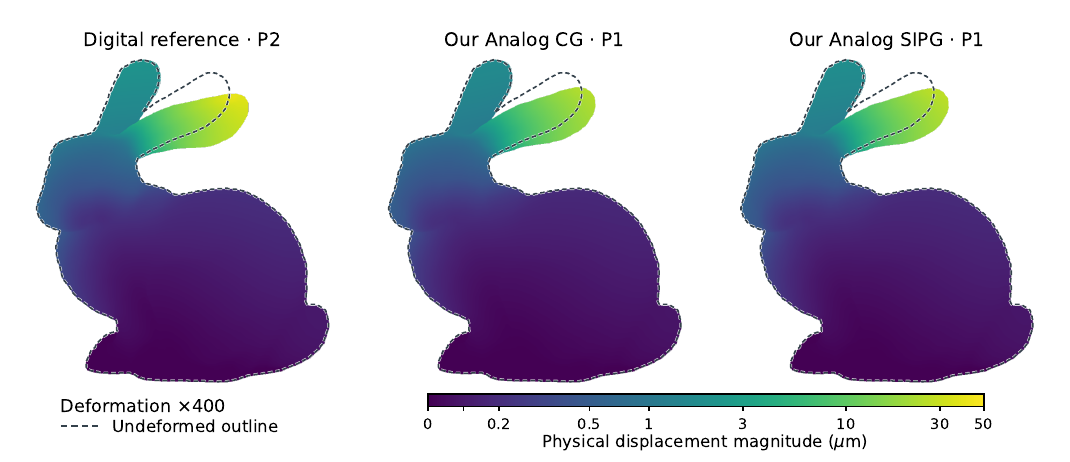}
\caption{Ear deflection at $400\times$ magnification. Dashed lines show the
undeformed outline. Color gives physical displacement in micrometres.}
\label{fig:bunny-nodal-warp}
\end{figure}

\begin{figure}[!htbp]\centering
\includegraphics[width=\linewidth]{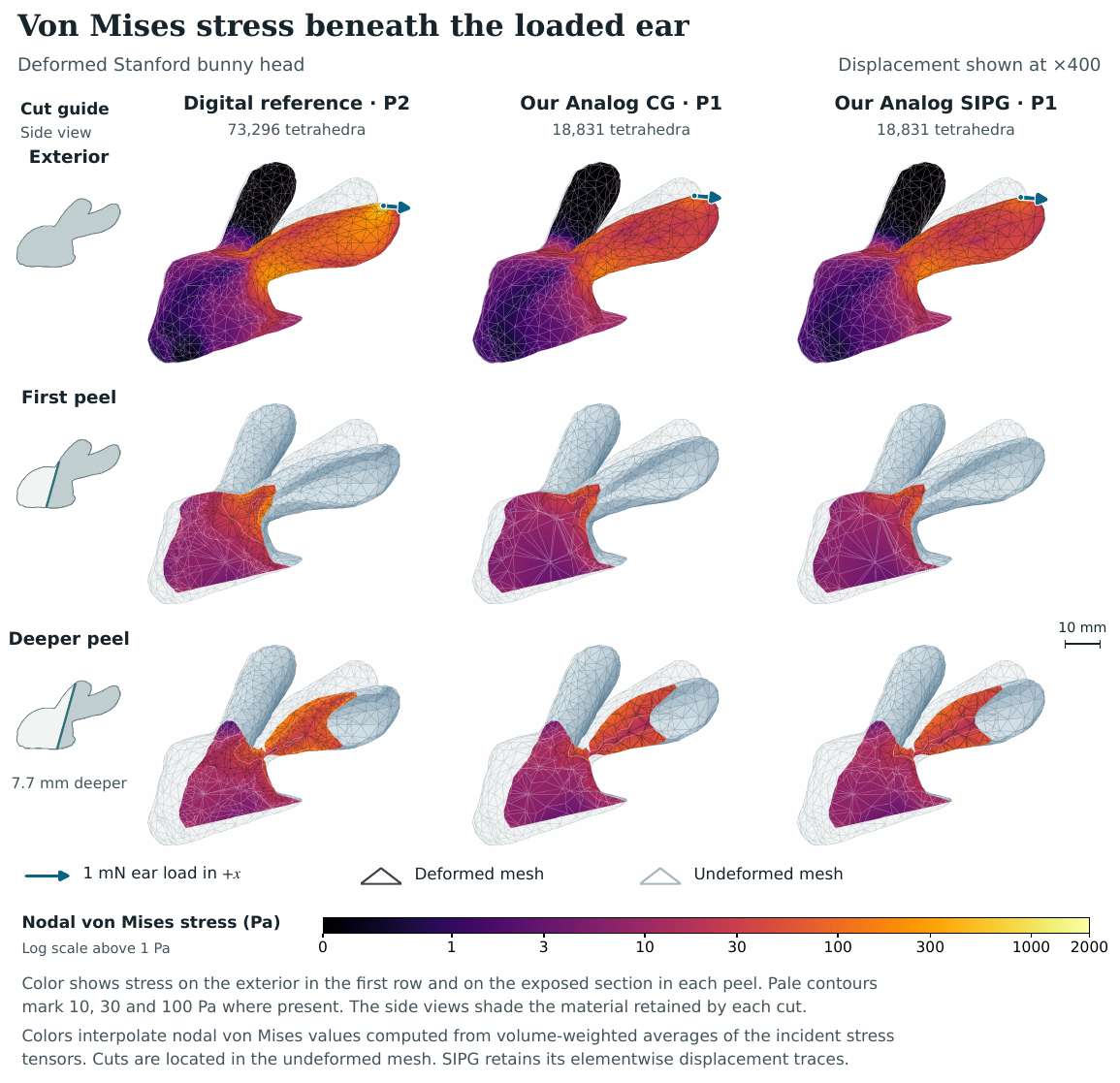}
\caption{Nodal von Mises stress on the exterior and two cuts spaced
7.7~mm apart in the undeformed head. Deformation is magnified $400\times$.
The faint mesh shows the original shape and arrows mark the 1~mN ear load.}
\label{fig:bunny-head-peel}
\end{figure}

In Figure~\ref{fig:bunny-nodal-warp} the loaded ear bends visibly away from its
original outline, and Figure~\ref{fig:bunny-head-peel} follows the stress into
the ear on parallel sections through the deformed head. To draw smooth contours
on these sections, we average the incident element stress tensors at each
vertex with tetrahedron volumes as weights, compute the nodal von Mises stress,
and interpolate it across the cut. This averaging affects only the display, and
all field comparisons use the original element values.

%% file: references.bib
@phdthesis{macneal1949,
  author = {Richard Henri MacNeal},
  title = {The Solution of Partial Differential Equations by Means of Electrical Networks},
  school = {California Institute of Technology},
  year = {1949},
  doi = {10.7907/PZ04-5290},
  url = {https://thesis.caltech.edu/1529/},
  note = {\doi{10.7907/PZ04-5290}}
}

@article{zidan2018,
  author = {Mohammed A. Zidan and YeonJoo Jeong and Jihang Lee and Bing Chen and Shuo Huang and Mark J. Kushner and Wei D. Lu},
  title = {A General Memristor-based Partial Differential Equation Solver},
  journal = {Nature Electronics},
  volume = {1}, number = {7}, pages = {411--420}, year = {2018},
  doi = {10.1038/s41928-018-0100-6}
}

@article{barnoski1966,
  author = {R. L. Barnoski and C. R. Freberg},
  title = {Passive-Element Analog Circuits for Three-Dimensional Elasticity},
  journal = {Journal of Engineering for Industry},
  volume = {88}, number = {1}, pages = {8--16}, year = {1966},
  doi = {10.1115/1.3670900}
}

@article{save2011,
  author = {Yogesh Dilip Save and H. Narayanan and Sachin B. Patkar},
  title = {Solution of Partial Differential Equations by Electrical Analogy},
  journal = {Journal of Computational Science},
  volume = {2}, number = {1}, pages = {18--30}, year = {2011},
  doi = {10.1016/j.jocs.2010.12.006}
}

@article{li2025iteration,
  author = {Jiancong Li and Yibai Xue and Yi Li and Han Jia and Zhiwei Zhou and Ling Yang and Shengguang Ren and Jia Chen and Yuhui He and Kanhao Xue and Ming Xu and Xiangshui Miao},
  title = {Fully Analog Iteration for Solving Matrix Equations with In-memory Computing},
  journal = {Science Advances}, volume = {11}, number = {7},
  pages = {eadr6391}, year = {2025},
  doi = {10.1126/sciadv.adr6391},
  url = {https://pmc.ncbi.nlm.nih.gov/articles/PMC11817932/}
}

@article{zuo2025,
  author = {Pushen Zuo and Qishen Wang and Yubiao Luo and Ruiqing Xie and Shiqing Wang and Zezhi Cheng and Lin Bao and Zongwei Wang and Yimao Cai and Ru Huang and Zhong Sun},
  title = {Precise and Scalable Analogue Matrix Equation Solving Using Resistive Random-access Memory Chips},
  journal = {Nature Electronics}, volume = {8}, number = {12}, pages = {1222--1233},
  year = {2025}, doi = {10.1038/s41928-025-01477-0},
  url = {https://www.nature.com/articles/s41928-025-01477-0}
}

@manual{tetgenmanual,
  author = {Hang Si},
  title = {{TetGen}, Version 1.5, User's Manual},
  organization = {Weierstrass Institute for Applied Analysis and Stochastics},
  year = {2013},
  note = {Also available as WIAS Technical Report No.~13, \doi{10.20347/WIAS.TECHREPORT.13}. Mesh generation, refinement, file formats and quadratic output},
  url = {https://wias-berlin.de/software/tetgen/1.5/doc/manual/manual.html}
}

@inproceedings{jiang2014rram,
  author = {Zizhen Jiang and Shimeng Yu and Yi Wu and Jesse H. Engel and Ximeng Guan and H.-S. Philip Wong},
  title = {{Verilog-A} Compact Model for Oxide-based Resistive Random Access Memory ({RRAM})},
  booktitle = {2014 International Conference on Simulation of Semiconductor Processes and Devices (SISPAD)},
  year = {2014},
  pages = {41--44},
  doi = {10.1109/SISPAD.2014.6931558},
  url = {https://in4.iue.tuwien.ac.at/pdfs/sispad2014/SISPAD_2014_041-044.pdf}
}

@manual{stanfordrramguide,
  author = {Zizhen Jiang and Shimeng Yu and Ximeng Guan and H.-S. Philip Wong},
  title = {A Quick User Guide on {Stanford University} Resistive-Switching Random Access Memory ({RRAM}) {Verilog-A} Model v.1.0.0},
  organization = {Stanford University},
  year = {2014},
  month = may,
  note = {Manual in the nanoHUB distribution of the model, \doi{10.4231/D37H1DN48}. Table 3 gives the default parameters and gap limits},
  url = {https://nanohub.org/publications/19/1}
}

@misc{ngsolvedg,
  author = {{NGSolve developers}},
  year = {n.d.},
  title = {Discontinuous {Galerkin} Methods},
  howpublished = {Official Netgen/NGSolve tutorial, Unit 2.8},
  url = {https://ngsolve.org/ngsolve/docs/i-tutorials/unit-2.8-DG/DG.html},
  note = {Interior-penalty formulation. Accessed October 30, 2026}
}

@misc{dealiisips,
  author = {Katharina Kormann and Martin Kronbichler},
  year = {2022},
  title = {The step-59 Tutorial Program},
  howpublished = {deal.II 9.4.0 documentation on the symmetric interior penalty formulation for the {Laplacian}},
  url = {https://dealii.org/9.4.0/doxygen/deal.II/step_59.html},
  note = {Accessed October 30, 2026}
}

@misc{gf180models,
  author = {{GlobalFoundries PDK Authors}},
  title = {{FAB3E} {Spice} Models for 0.18\,$\mu$m {3.3V/6V} High Voltage {MCU} Process: Model Reference Guide},
  howpublished = {{GlobalFoundries} {GF180MCU} {PDK} documentation},
  year = {2022},
  url = {https://gf180mcu-pdk.readthedocs.io/en/latest/analog/model_parameters/LV/LV.html},
  note = {Section~1.4.1 lists nfet\_06v0 and pfet\_06v0 as {BSIM4} models for 6~V {MOSFETs}, and Section~2 gives the {MOSFET} models and process corners. Accessed November 26, 2026}
}

@article{tianstability,
  author = {Michael Tian and V. Visvanathan and Jeffrey Hantgan and Kenneth Kundert},
  title = {Striving for Small-Signal Stability},
  journal = {IEEE Circuits and Devices Magazine},
  year = {2001},
  volume = {17},
  number = {1},
  pages = {31--41},
  doi = {10.1109/101.900125},
  url = {https://kenkundert.com/docs/cd2001-01.pdf},
  note = {Loop-based two-port and device-based stability algorithms}
}

@misc{stanfordbunny,
  author = {{Stanford University Computer Graphics Laboratory}},
  title = {{Stanford Bunny}},
  year = {1994},
  howpublished = {The Stanford 3D Scanning Repository},
  url = {https://graphics.stanford.edu/data/3Dscanrep/}
}
